\def\virg#1{``#1''}

\def\rfr#1{eq. (\ref{#1})}

\def\derp#1#2{\rp{\partial{#1}}{\partial{#2}}}
\def\dert#1#2{\frac{{{d}}{#1}}{{{d}}{#2}}}              % derivate parziali e totali prima e seconda

\def\bar{\begin{eqnarray}}
\def\ear{\end{eqnarray}}
\def\bb{\bibitem}
\def\eqI{\begin{equation}}
\def\eqF{\end{equation}}
\def\eqia{\begin{eqnarray}}
\def\eqfa{\end{eqnarray}}
\def\rp#1#2{{#1\over#2}}

\def\lb#1{\label{#1}}

\def\oc2{$\mathcal{O}(c^{-2})$}

\documentclass{aastex}
\usepackage{url}
\usepackage{amsmath,amsthm,amscd,amssymb}
\usepackage{latexsym}
\usepackage{graphicx,epsfig}

\RequirePackage{color}

\begin{document}

\title{On the impact of the atmospheric drag on the LARES mission}

\shorttitle{On the impact of the atmospheric drag on the LARES mission}
\shortauthors{L. Iorio }

\author{Lorenzo Iorio }
\affil{INFN-Sezione di Pisa. Permanent address for correspondence: Viale Unit\`{a} di Italia 68, 70125, Bari (BA), Italy. E-mail: lorenzo.iorio@libero.it}

\begin{abstract}
 The goal of the recently approved space-based LARES mission is to measure the general relativistic Lense-Thirring effect in the gravitational field of the spinning Earth at a repeatedly claimed $\approx 1\%$ accuracy by combining its node $\Omega$ with those of the existing LAGEOS and LAGEOS II laser-ranged satellites. In this paper we show that, in view of the lower altitude of LARES ($h=1450$ km) with respect to LAGEOS and LAGEOS II ($h\simeq 6000$ km), the cross-coupling between the effect of the atmospheric drag, both neutral and charged, on the inclination of LARES and its classical node precession due to the Earth's oblateness may induce a $3-9\%$ year$^{-1}$  systematic bias on the total relativistic precession.  Since its extraction from the data will take about $5-10$ years, such a perturbing effect may degrade the total accuracy of the test, especially in view of the large uncertainties in modeling the drag force.
 \end{abstract}


%\pacs{04.80.Cc 91.10.Sp 94.05.Hk 91.10.Qm}
\section{Introduction}
The LARES (LAser RElativity Satellite) satellite, recently approved\footnote{See on the WEB http://www.asi.it/SiteEN/MotorSearchFullText.aspx?keyw=LARES} by the Italian Space Agency, should have been launched at the end of\footnote{See on the WEB
http://www.esa.int/esapub/bulletin/bulletin135/bul135f$\_$bianchi.pdf. Actually, the launch date has been postponed to late 2010/early 2011. http://www.spacenews.com/civil/100115-asi-expects-budget-remain-flat-2010.html} 2009 with a VEGA rocket in a circular orbit inclined by $71.5$ deg to the Earth's equator at an
altitude of\footnote{In its originally proposed configuration \citep{Ciu86} the semi-major axis of LARES was equal to that of LAGEOS, i.e. $a=12270$ km.} 1450 km \citep{ESRIN}.  Its goal is a measurement of the general relativistic gravitomagnetic Lense-Thirring  effect \citep{LT} due to the Earth's rotation  at a repeatedly claimed $\simeq 1\%$ level of accuracy in conjunction with the existing LAGEOS and LAGEOS II laser-ranged satellites which fly at much higher altitudes, i.e. $h\simeq 6000$ km. The observable is a suitable linear combination of the longitudes of the ascending nodes $\Omega$ of the three satellites, because the gravitomagnetic field of the Earth induces   a secular precession   on such a Keplerian orbital element
\eqI\dot\Omega_{\rm LT}=\rp{2GL}{c^2 a^3(1-e^2)^{3/2}},\eqF
where $G$ is the Newtonian gravitational constant, $L$ is the Earth's spin angular momentum, $c$ is the speed of light in vacuum, $a$ is the satellite's semi-major axis and $e$ is its eccentricity.
In Table \ref{OSIGNUR} we quote the Lense-Thirring precessions for LAGEOS, LAGEOS II and LARES: their magnitudes are of the order of $10^1-10^2$ milliarcseconds per year (mas yr$^{-1}$ in the following).
\begin{table}[!ht]
\caption{LAGEOS, LAGEOS II and LARES: orbital parameters and node precessions due to the terrestrial gravitomagnetic field and the first two even zonal harmonics  for
$L_{\oplus} = 5.86\times 10^{33}$ kg m$^2$ s$^{-1}$ \protect\citep{IERS}, $J_2=0.00108263538$, $J_4=-1.619989\times 10^{-6}$. \citep{ggm03}}\label{OSIGNUR}
%\centering
%
%\bigskip
\begin{tabular}{lllllll}
\hline\noalign{\smallskip}
Satellite & $a$ (km) & $e$ & $I$ (deg) & $\dot\Omega_{\rm LT}$ (mas yr$^{-1}$) & $\dot\Omega_{J_{2}}$ (mas yr$^{-1}$)& $\dot\Omega_{J_{4}}$ (mas yr$^{-1}$)\\
\noalign{\smallskip}\hline\noalign{\smallskip}
LAGEOS & 12270 &  0.0045 & 109.9 & 30.7 & $4.538082658\times 10^{8}$ & $-2.501490\times 10^{5}$ \\
LAGEOS II & 12163 &  0.014 & 52.65 & 31.5 & $-8.303252509\times 10^{8}$ & $9.05051\times 10^{4}$\\
LARES & 7828 &  0.0 & 71.5 & 118.1 & $-2.0298207310\times 10^{9}$ & $2.8925357\times 10^{6}$ \\
\noalign{\smallskip}\hline\noalign{\smallskip}
\end{tabular}
\end{table}

The  much larger classical secular precessions  induced  on $\Omega$ by the even zonal harmonic coefficients $J_{\ell},\ \ell=2,4,6,...$  of the multipolar expansion of the terrestrial gravitational potential accounting for the centrifugal oblateness of our planet \citep{Kau} are a major source of systematic uncertainty. They can be written as
\eqI \dot\Omega^{\rm obl}=\sum_{\ell=2}\dot\Omega_{.\ell}J_{\ell},\eqF
where the coefficients $\dot\Omega_{.\ell}$ depend on the Earth's mass $M$ and equatorial radius $R$, and of the orbital geometry of the satellite through $a$, $e$ and the inclination $I$ of the orbital plane to the Earth's equator.
Since they have the same temporal signature of the relativistic effect of interest, they cannot be subtracted from the signal without affecting the recovery of the Lense-Thirring effect itself. Thus, it is of the utmost importance to realistically assess the uncertainty in them in order to evaluate their percent impact on the gravitomagnetic shift. More specifically, the magnitude of the node secular precessions due to the first two even zonals for LAGEOS, LAGEOS II and LARES are listed in Table \ref{OSIGNUR}. It can be noted that the $J_2-$induced rates are of the order of $10^{8}-10^{9}$ mas yr$^{-1}$, i.e. seven orders of magnitude larger than the Lense-Thirring precessions. The perturbations by $J_4$ are of the order of $10^{5}-10^{6}$ mas yr$^{-1}$, i.e. four orders of magnitude larger than the gravitomagnetic effects. Such figures immediately demonstrate the difficulty of determining a smallish relativistic effect with respect to a huge classical one, which needs, thus, to be accounted for with the appropriate accuracy (about one part in $10^{10}$), or, as combining the three satellite data sets aims at (see Section \ref{secunda}), removed from the signal with the same accuracy.

Up to now major efforts  have been devoted to evaluate the bias due to the lingering uncertainty $\delta J_{\ell}$ in the even zonals according to
\eqI\delta\dot\Omega^{\rm obl}_{J_{\ell}} \leq \sum_{\ell=2}|\dot\Omega_{.\ell}|\delta J_{\ell}.\eqF A reliable evaluation of such a corrupting effect is made difficult by the fact that the relatively low altitude of LARES brings into play more even zonals than done by LAGEOS and LAGEOS II \citep{Ior1}.

Concerning the non-conservative orbital perturbations \citep{Mil} like direct solar radiation pressure, Earth's albedo, direct Earth's infrared radiation,
atmospheric drag, thermal effects like the Yarkovski-Schach and Rubincam ones, they have been so far regarded as a minor concern because their direct impact on the node of the LAGEOS-type satellites is  $\lesssim 1\%$ of the Lense-Thirring effect \citep{Luc1,Luc}.

In this paper we want to investigate their indirect effects through
the cross-coupling  \citep{Kau}
\eqI\delta\dot\Omega_{I}^{\rm obl}\leq \left|\derp{\dot\Omega_{.\ell} }{I} J_{\ell} \right|\delta I\eqF
between the zonals-induced node precessions and certain non-gravitational perturbations  affecting the LARES inclination.
We will show that, in particular, the impact of the atmospheric drag on $I_{\rm LR}$ may play an important role in the evaluation of the error budget of the Lense-Thirring test. Indeed, although the direct secular effect of the atmospheric drag on the node vanishes, it is not so for the indirect one due to the non-vanishing secular decrease of the inclination which maps onto a node effect. Moreover, we will point out that it should not be possible to correct the signal for the measured value of $I$ arc by arc without likely affecting the gravitomagnetic signal of interest itself. Clearly, since the non-gravitational perturbations do depend on the particular type of satellite considered and since LARES has not yet been launched in orbit, our investigation should not be required to be more accurate than it can be in the sense that it must be viewed as a reasonable sensitivity analysis pointing out a possible source of potential bias and evaluating conservatively the largest possible size of the effect examined.

The paper is organized as follows. In Section \ref{primera} we calculate the secular rate of the inclination of a LAGEOS-type satellite induced by a drag force and compute it for LARES. In Section \ref{secunda} we calculate its indirect effect on the Lense-Thirring shift through the secular precession due to the even zonal harmonics. We also briefly discuss other non-gravitational perturbations which may cause a secular variation of the LARES inclination in Section \ref{tria}. Section \ref{quarta} is devoted to the conclusions.

\section{The effect of the atmospheric drag on the inclination of LARES}\lb{primera}
The Gauss equation for the variation of the inclination $I$ is  \citep{Mil}
\eqI \dert I t = \rp{r\cos u}{na^2\sqrt{1-e^2}}A_{\nu},\lb{gaus}\eqF
where $n\doteq\sqrt{GM/a^3}$ is the un-perturbed Keplerian mean motion, $u \doteq g+f$ is the argument of latitude, defined as the sum of the argument of pericentre $g$, which fixes the position of the pericentre with respect to the line of the nodes, and the true anomaly $f$ which reckons the instantaneous position of the spacecraft from the pericentre, and $A_{\nu}$ is the out-of-plane component of the perturbing acceleration $\boldsymbol{A}$.

The drag force per unit mass is  \citep{King}
\eqI
\boldsymbol{A}_{\rm D} = -\rp{1}{2}C_D\Sigma\rho V\boldsymbol{V}\lb{atmo},
\eqF
where  \eqI\boldsymbol{V}=\boldsymbol{v}-\boldsymbol{V}_{\rm A}\eqF is the satellite velocity with respect to the atmosphere; $\boldsymbol{v}$ and $\boldsymbol{V}_{\rm A}$ are the geocentric satellite and atmosphere velocities, respectively. %and\footnote{It holds for small eccentricities.}
%\eqI Z\simeq \left(1-\rp{a\omega_{\rm A}}{v}\cos i\right)^2,\eqF
%in which $\omega_{\rm A}$ is the atmosphere angular speed; it can be assumed there is a $\simeq 20\%$ uncertainty in it, so that it can be posed
%\eqI\omega_{\rm A}\simeq(1+0.2)\omega_{\oplus}.\eqF
The other parameters entering \rfr{atmo}
are the drag coefficient $C_D$, which depends in a complicated way on the interaction between the gas of particles in the surroundings of the satellite and its surface \citep{Afo85,Mil}, $\Sigma\doteq S/m$ is the area-to-mass ratio\footnote{$S$ denotes the spacecraft cross sectional area (perpendicular to the velocity).}  of the satellite, and $\rho$ is the density of the atmosphere.

The velocity of the atmosphere,  known as ambient velocity, can be written in terms of geocentric inertial quantities as
\eqI\boldsymbol{V}_{\rm A} = \boldsymbol{\omega}_{\rm A}\boldsymbol{\times}\boldsymbol{r},\eqF
with \eqI\boldsymbol{\omega}_{\rm A}=(1+\xi)\boldsymbol{\omega}_{\oplus}=(1+\xi)\omega_{\oplus}\ \boldsymbol{k},\lb{rot}\eqF
where $\boldsymbol{k}$ is the unit vector of the $z-$axis in an inertial geocentric frame chosen aligned with the Earth's angular velocity vector.
Note that \rfr{rot} accounts for the fact that the atmosphere, in general, does not co-rotate exactly with the Earth; maximum  observed deviations from the simplifying assumption of exact co-rotation are of the order of $40\%$ \citep{King}.
Thus,
\eqI\boldsymbol{V}_{\rm A}= \omega_{\rm A}\left(-y\ \boldsymbol{i} + x\ \boldsymbol{j}\right),\lb{vatmo}\eqF
where $\boldsymbol{i}$ and $\boldsymbol{j}$ are the unit vectors in the reference $\{xy\}$ plane of the geocentric inertial frame which coincides with the Earth's equator; the angle between $\boldsymbol{v}$, which lies in the orbital plane, and $\boldsymbol{V}_{\rm A}$ is the inclination $I$.

In order to have the out-of-plane component $A_{\nu}$ of the drag acceleration evaluated onto the un-perturbed Keplerian ellipse,
to be inserted into the right-hand-side of \rfr{gaus}, $\boldsymbol V$ must be projected onto the $\boldsymbol{\hat{n}}$ direction
of the frame co-moving with the satellite; since
\eqI\boldsymbol{\hat{n}}=\sin I\sin\Omega\ \boldsymbol{i} -\sin I\cos\Omega\ \boldsymbol{j}+\cos I\ \boldsymbol{k},\lb{kappa}\eqF
then, by choosing $\Omega =0$, \eqI \boldsymbol{V}_{\rm A}\cdot\boldsymbol{\hat{n}} = -\omega_{\rm A} x\sin I.\eqF
Onto the unperturbed orbit
\eqI x=r\cos u\cos\Omega - \sin u\cos I\sin \Omega,\eqF
%\eqI y = r\cos u\sin\Omega + \sin u\cos i\cos \Omega,\eqF
%\eqI z = r\sin u\sin i,\eqF
so that it is possible to obtain
 \eqI \boldsymbol{V}_{\rm A}\cdot\boldsymbol{\hat{n}} = -\omega_{\rm A} r\sin I\cos u.\eqF
Since \eqI\boldsymbol{v}=v_r\ \boldsymbol{\hat{r}}+v_t\ \boldsymbol{\hat{t}},\eqF it appears clear that if the Earth's atmosphere did not rotate there would not be any out-of-plane component of the drag acceleration which, instead, exists because  $\boldsymbol{V}_{\rm A}\cdot \boldsymbol{\hat{n}}\neq 0$ for non-equatorial orbits.
Thus, the out-of-plane component of \rfr{atmo} is
\eqI A_{\nu} = -\rp{1}{2}C_D\Sigma\rho V \omega_{\rm A} r\sin I\cos u.\eqF
%Concerning $V$ , it can be obtained by noting that the radial and transverse unit vectors of the co-moving frame can be written as\footnote{As before, %we choose $\Omega = 0$.}
%\eqI \boldsymbol{\hat{r}} = \cos u\ \boldsymbol{i} + \sin u\cos i\ \boldsymbol{j} +\sin u\sin i\ \ \boldsymbol{k}, \eqF
% \eqI \boldsymbol{\hat{t}} = -\sin u\ \boldsymbol{i} + \cos u\cos i\ \boldsymbol{j} +\cos u\sin i\ \ \boldsymbol{k}, \eqF
It turns out that it can be posed  \citep{SALAM}
\eqI V=|\boldsymbol{v}-\boldsymbol{V}_{\rm A}|\simeq v\sqrt{k_{\rm R}}, \ k_{\rm R}\simeq 1,\lb{sala}\eqF
so that
\eqI A_{\nu}\simeq -\rp{1}{2}C_D\Sigma\rho v \omega_{\rm A} r\sin I\cos u.\lb{accia}\eqF
Concerning the approximations used in \rfr{sala} and \rfr{accia}, they are justified since
\eqI k_{\rm R}\doteq 1+\left(\rp{V_{\rm A}}{v}\right)^2 -2\left(\rp{V_{\rm A}}{v}\right)\cos I, \eqF
where typically $V_{\rm A}\simeq 0.5$ km s$^{-1}$ because of \rfr{rot} and \rfr{vatmo} (see also Table \ref{tavola1}), and $v\simeq \sqrt{GM/a}=7.1$ km s$^{-1}$ for orbital heights of about 1400 km.

By inserting \rfr{accia} into \rfr{gaus} with  the un-perturbed relations
\eqI r = \rp{a(1-e^2)}{1+e\cos f},\ v=na\sqrt{ \rp{1+e^2+2e\cos f}{1-e^2}  },\eqF
 and integrating over an orbital period $P_{\rm b}$ by means of
\eqI \rp{dt}{P_{\rm b}} = \rp{(1-e^2)^{3/2}}{2\pi (1+e\cos f)^2}df,\eqF
one finds that there is a non-vanishing secular rate of the inclination to order zero in the eccentricity \eqI \left\langle\dert I t\right\rangle = -\rp{1}{4}C_D\Sigma\rho\omega_{\rm A}a\sin I;\lb{didt}\eqF it agrees with (6.17) by\footnote{$\Delta I$ in \citep{Mil} is the shift per revolution; in order to be confronted with \rfr{didt}, (6.17) by \citet{Mil} must be divided by $P_{\rm b}=2\pi/n$. By putting $Z\rightarrow 1$ and $v=na$ one recovers just \rfr{didt}.} \citet{Mil}.
In obtaining \rfr{didt} we considered the atmospheric density $\rho$ constant over one orbital revolution; since for LARES $P_{\rm b}=3.7$ h, this is certainly a reasonable assumption. In general, $\rho$ undergoes many irregular and complex variations both in position and time, being largely affected by solar activity and by the heating and cooling of the atmosphere \citep{King,SALAM}.
\begin{table}[!ht]
\caption{Relevant physical and orbital parameters of the Earth-LARES system. The quoted value for $C_D$ is usually used in literature, but it refers typically to altitudes of some hundreds km; at 1450 km it may be larger \citep{Mil}. The value of the area-to-mass ratio $\Sigma$ has been obtained by using for LARES a diameter of $d=37.6$ cm and a mass of $m=400$ kg (http://esamultimedia.esa.int/docs/LEX-EC/CubeSat$\%$20CFP$\%$20issue$\%$201.pdf). The value of $\rho$ is that for the Ajisai satellite \citep{Aji} which has a semimajor axis of 7870 km. Concerning the rotation of the atmosphere, the quoted value has been obtained by assuming it is about 20$\%$ faster than the Earth itself.}
\begin{tabular}{llll}
\hline\noalign{\smallskip}
Parameter & Value & Units & Reference \\
\noalign{\smallskip}\hline\noalign{\smallskip}
$GM_{\oplus}$ & $3.986004418\times 10^{14}$ & m$^3$ s$^{-2}$ & \citep{IERS}\\
$R_{\oplus}$ & $6378136.6$ & m & \citep{IERS}\\
$J_2$ & $0.00108263538$ &- & \citep{ggm03}\\
$a_{\rm LR}$ & $7828\times 10^3$ & m & \citep{ESRIN}\\
$e_{\rm LR}$ & 0 & -& \citep{ESRIN}\\
$I_{\rm LR}$ & 71.5& deg & \citep{ESRIN}\\
$C_D$ & $2.2$ & - & \citep{SALAM}\\
$\Sigma$& $3\times 10^{-4}$ & m$^2$ kg$^{-1}$ & (See\ caption)\\
$\rho$ & $1\times 10^{-15}$ & kg m$^{-3}$ & \citep{Aji}\\
$\omega_{\rm A}$ & $8.750538\times 10^{-5}$ & s$^{-1}$ & -\\
\noalign{\smallskip}\hline\noalign{\smallskip}
\end{tabular}
\end{table}

According to Table \ref{tavola1}, the inclination of LARES will experience a secular decrease of
 \eqI \left\langle\dert I t\right\rangle_{\rm LR} =-3\times 10^{-9}\ {\rm rad}\ {\rm yr}^{-1} = -0.6\ {\rm mas}\ {\rm yr}^{-1},\eqF
where mas stands for milli-arcseconds. Concerning the node, whose Gauss variation equation is identical to \rfr{gaus} with $\cos u$ replaced by $\sin u/\sin I$, it can be shown that there are no secular effects induced by the atmospheric drag on it; the first non-vanishing term is proportional to $e^2\sin 2g$.

\section{The impact of the secular decrease of the inclination on the node precession due to the oblateness}\lb{secunda}
Such a decrease of $I_{\rm LR}$ affects also the secular precession of the spacecraft node due to the oblateness of the Earth which is a major corrupting effect for the Lense-Thirring signal.

Indeed, since
\eqI \dot\Omega_{J_2} = -\rp{3}{2}n\left(\rp{R}{a}\right)^2\rp{\cos I\ J_2}{(1-e^2)^2}\eqF
a bias
\eqI \delta\dot\Omega_{i} = \rp{3}{2}n\left(\rp{R}{a}\right)^2\rp{\sin I\ J_2}{(1-e^2)^2}\left\langle \dert I t\right\rangle \Delta t\lb{culo}\eqF
occurs. For LARES \rfr{culo} yields a shift of $18.8$ mas yr$^{-1}$ over one year; since the Lense-Thirring precession of the node of LARES amounts to
118 mas yr$^{-1}$, the cross-coupling of the inclination perturbation with the oblateness would yield a systematic error of $16\%$ over just one year.

In fact, the data of LARES will be combined with those of the existing LAGEOS and LAGEOS II spacecraft according to the following linear combination of their nodes \citep{IorNA}
\eqI \dot\Omega^{\rm LAGEOS} + c_1\dot\Omega^{\rm LAGEOS\ II} +c_2\dot\Omega^{\rm LARES},\ c_1=0.358642219,\ c_2=0.075117522\lb{comb}\eqF   in order to cancel out the impact of the mismodelling $\delta J_2$ and $\delta J_4$ of the first two even zonal harmonics; the general relativistic prediction of the total Lense-Thirring shift, according to the linear combination of \rfr{comb}, is 50.7  mas yr$^{-1}$.  The combination of \rfr{comb} is based on a strategy put forth for the first time in \citep{Ciu96}. It turns out that the impact of \rfr{didt} on \rfr{comb} is $3\%$ yr$^{-1}$. Note that since $c_1$ and $c_2$ are aimed at removing the classical effects by $J_2$ and $J_4$ to the needed extent, we released them with nine decimal digits, i.e. with the matching accuracy. Indeed, as shown by Table \ref{OSIGNUR}, the largest classical effect is about seven orders of magnitude larger than the Lense-Thirring precessions, and the claimed accuracy of the proposed test is $1\%$.
In obtaining such a result we treated the coefficients $c_1$ and $c_2$, which depend on the semi-major axes, the eccentricities and the inclinations of the three satellites, as  constant numbers; let us check if a conservative uncertainty of the order of\footnote{In fact, it may likely be about one order of magnitude smaller because $\delta r\simeq 1$ cm yields an uncertainty of approximately 0.3 mas at an height of 1450 km.} $\delta I_{\rm LR}\simeq 1$ mas in $I_{\rm LR}$ can affect the numerical values of $c_1$ and $c_2$ at the ninth decimal digit. It turns out that
\eqI \left| c_1(I_{\rm LR})-c_{1}(I_{\rm LR}+\delta I_{\rm LR})\right|=1.7\times 10^{-9}, \left| c_2(I_{\rm LR})-c_{2}(I_{\rm LR}+\delta I_{\rm LR})\right|=3\times 10^{-10}.\eqF Thus, we can conclude that the uncertainty in determining the inclination of LARES is not a concern about the accuracy required to compute $c_1$ and $c_2$. By repeating the same analysis for the semimajor axis $a_{\rm LR}$ of LARES, it turns out that an uncertainty $\delta a_{\rm LR}\simeq 1$ cm yields a similar conclusion because
\eqI \left| c_1(a_{\rm LR})-c_{1}(a_{\rm LR}+\delta a_{\rm LR})\right|=5\times 10^{-10}, \left| c_2(a_{\rm LR})-c_{2}(a_{\rm LR}+\delta a_{\rm LR})\right|=5\times 10^{-10}.\eqF

%indeed, it turns out that its dependence on $I_{\rm LR}$ is completely negligible for variations of $I_{\rm LR}$ of the order of few mas. The same holds also %for $c_1$.

Let us see what could be the impact of the uncertainties in parameters like $C_D$ and $\omega_{\rm A}$ on our estimates. For $2< C_D < 2.5$ we get a substantially unchanged bias $2.7-3.4\%$ yr$^{-1}$. By assuming $\omega_{\rm A}=\omega_{\oplus} = 7.292115\times 10^{-5}$ s$^{-1}$, i.e. by assuming that the atmosphere co-rotates with the Earth, the bias amounts to $2.5\%$ yr$^{-1}$.  Concerning the approximation of \rfr{sala} used for $V$, i.e. $V=v\sqrt{k_{\rm R}}\simeq v$, it is fully justified in our case. Indeed, $(V_{\rm A}/v)\cos I$ appearing in it can be approximated with $(\omega_{\rm A}/n)\cos I$ for a circular orbit; for LARES it amounts to 0.03 only, thus yielding $\sqrt{k_{\rm R}}=0.97$. It must be noted that the effect of \rfr{culo} should likely affect the  LARES data in full because of the difficulty of realistically modelling the drag force, especially $C_D$ and $\rho$; just to give an idea of the uncertainty in their values note that when the solar activity is low a typical atmospheric density at about 1500 km altitude is $2\times 10^{-16}$ kg m$^{-3}$ \citep{Aji}, while for the existing LAGEOS satellite the drag coefficient is $C_D\simeq 4.9$ \citep{Afo85,Mil}. The fact that $C_D$ is larger for LAGEOS than for a lower satellite like LARES is only seemingly contradictory (the higher the orbit, the smaller the drag). Indeed, the drag coefficient depends on the ratio between the average thermal molecular speed of the atmosphere $V_{\rm T}$ and the orbital speed $v$ of the spacecraft. For relatively low orbits, $V_{\rm T}\simeq 1$ km s$^{-1}$ is typically smaller than\footnote{For LARES (see Table \ref{OSIGNUR} for its orbital parameters) it is $v=7.1$ km s$^{-1}$.} $v\simeq 7.5$ km s$^{-1}$. On the contrary, at higher altitudes the situation changes because $v$ becomes smaller and smaller, while $V_{\rm T}$ increases rapidly due to higher temperatures and lower mean molecular weight. In this case, as for LAGEOS, \citep{Afo85}
\eqI C_D = b\left[2+\rp{4}{3}\left\langle\left(\rp{V_{\rm T}}{v}\right)^2\right\rangle-\rp{2}{15}\left\langle\left(\rp{V_{\rm T}}{v}\right)^4\right\rangle\right],\eqF with $b\simeq 1.4$ and $\left\langle V_{\rm T}/v\right\rangle\simeq 0.8$, so that $C_D\simeq 4$. For more details, see the discussion in \citep{Mil}, pp.104-107.

In addition to the neutral particle drag considered so far it should also be taken into account the charged particle drag \citep{Afo85} due to the fact that a spacecraft moving in a gas of electrons and ions tends to acquire an electric charge because of the collisions with such particles and also because of the photoelectric effect caused by solar radiation. The effect of the charged particle drag can be obtained by re-scaling the one due to the neutral particle drag by a multiplicative factor $b$ containing, among other things, the satellite's potential $V_0$. According to \citet{Luc01}, it may amount to about $V_0=-0.3$ $V$ for LARES, so that $b=3.1$ which implies a $9\%$ yr$^{-1}$ systematic error in the measurement of the Lense-Thirring effect with \rfr{comb}.  It must be pointed out that the reduction of the impact of the perturbing accelerations of thermal origin should have  been reached by the LARES team with two concentric spheres. However, as explained by \citet{Andres}, this solution will increase the floating potential of LARES because of the much higher electrical resistivity, so that the evaluations presented here may turn out to be optimistic.

Since the extraction of the relativistic effect would require a multi-year analysis, typically $\Delta t = 5-10$ yr, the action of the overall atmospheric drag on the LARES inclination may be a serious corrupting effect over such timescales.

 \citet{megaciuf} objected that, in fact, the disturbing effect examined would not appear in the real data analysis procedure because the inclination along with all the other Keplerian orbital elements would be\footnote{Actually, the Keplerian orbital elements are not directly measurable quantities, contrary to, e.g., ranges, range-rates, right ascension, declination.} \virg{measured} arc by arc, so that one should only have to correct the signal for the measured value of the inclination; after all, the same problems, if not even larger, would occur with the semimajor axes of the LAGEOS satellites, which are known to undergo still unexplained secular decrease of 1.1 mm d$^{-1}$ \citep{Ruby} and their consequent mappings onto the node rates. The problem is that while a perturbation $\Delta a$ pertains the in-plane, radial $\mathcal{R}$ component \citep{Chri} of the LAGEOSs' orbits, both the Lense-Thirring node precession and the shifts in the inclination affect the out-of-plane, normal $\mathcal{N}$ component \citep{Chri} of the orbit; thus, even if repeated corrections to the semimajor axis could be applied without affecting the gravitomagnetic signal of interest, the same would not hold for the inclination. This is particularly true in view of the fact that, for still unexplained reasons, the Lense-Thirring effect itself has never been estimated, either as a short-arc or as a global parameter.
Moreover, \citet{megaciuf} claimed that the recent improvements in atmospheric refraction modelling would allow to \virg{measure} the inclination of the LAGEOSs satellites at
a level of accuracy, on average, of $30$ $\mu$as for LAGEOS  and $10$ $\mu$as for LAGEOS II. Firstly,  the
tracking of a relatively low satellite is always more difficult than for higher targets, so that
caution would be needed in straightforwardly extrapolating results valid for LAGEOS to the still non-existing LARES. Second,
it is difficult to understand the exact
sense of such claims because they would imply an accuracy $\delta r\simeq a\delta I$ in reconstructing the orbits of LAGEOS and LAGEOS II, on average, of
$0.2$ cm and $0.06$ cm, respectively.
\section{Other effects potentially inducing secular variations of the LARES inclination}\lb{tria}
Among the other non-conservative forces acting on the LAGEOS-type satellites, also the \citet{Rub} effect, due to the anisotropic re-emission of the infrared radiation of the Earth along the satellite's spin axis, is important. Such an effect arises from the fact that the retroreflectors of LAGEOS  have a significant thermal inertia of about 3000 s; since such time of thermal response is shorter than the satellite's orbital period $P_{\rm b} =13526$ s and larger than the  spin period, whose nominal value was about 1 s at the launch epoch, the perturbation induces a temperature asymmetry between the hemisphere facing the Earth and the one opposite to it.
The illuminated hemisphere becomes hotter than the dark one only after the Earth has passed its pole, causing a time lag effect accounted for by the thermal lag angle $\theta$.
A net recoil acceleration $A_{\rm Rub}$ directed along  the satellite spin axis occurs.  It induces a secular rate of the inclination according to
\citep{Luc}
\eqI\left\langle\dert I t\right\rangle = -\rp{A_{\rm Rub}}{8na}\sin\theta\sin 2 I\left(3\sigma_z^2 -1\right).\lb{rubi}\eqF
 In it $A_{\rm Rub}$ is the Rubincam acceleration which depends in a complex way on the physical and thermal properties of the satellite and of its array of retro-reflectors, $\theta$ is the thermal lag angle, and $\sigma_z$ is the component of the satellite's spin along the $z$ axis of a geocentric equatorial inertial frame having the $x$ axis along the vernal equinox direction. For LAGEOS II the secular inclination rate is of the order of 1.5 mas yr$^{-1}$. By assuming for LARES the same value of $A_{\rm Rub}$ as for LAGEOS II, i.e. $A_{\rm Rub}\simeq -7\times 10^{-12}$ m s$^{-2}$, \rfr{rubi} yields an effect of the order of about 0.7 mas yr$^{-1}$. In fact, it might be finally smaller because of the currently ongoing manufacturing efforts of the LARES team aimed at reducing the impact of the non-gravitational perturbations of thermal origin on the new spacecraft with respect to the LAGEOS satellites \citep{LAR}. Moreover, it will depend on the direction of the satellite's spin at the injection in orbit.

\section{Conclusions}\lb{quarta}
In this paper we have shown that certain subtle non-gravitational perturbations acting on the
forthcoming LARES satellite may corrupt the claimed goal of performing a $\simeq 1\%$ measurement of the
Lense-Thirring effect in the gravitational field of the rotating Earth because of the lower altitude of the new
spacecraft with respect to the existing LAGEOS and LAGEOS II spacecraft. In particular, the interplay between the node precessions due to the even zonal harmonics of the geopotential, which are a major source of systematic error, and the LARES inclination has been investigated. The atmospheric drag, both in its neutral and charged components, will induce a non-negligible secular decrease of the inclination of the new spacecraft yielding  a correction to the node precession of degree $\ell=2$  which amounts to $3-9\%$ yr$^{-1}$ of the total gravitomagnetic signal pertaining just the node itself. Such a corrupting bias would be very difficult to be modeled. Since the extraction of the relativistic signature will require a data analysis of about $5-10$ yr, the effect examined here may yield a degradation of the achievable total accuracy of the test. In principle, also the Rubincam effect, of thermal origin, should be taken into account because it can induce a non-vanishing secular variation of the inclination. Since both the node and the inclination enter the out-of-plane, normal component of the orbit of a satellite, it would not be possible to correct for the measured values of the inclinations arc by arc without likely affecting also the Lense-Thirring signal itself, especially because it has never been estimated along with the other parameters.

\section*{Acknowledgments}
I am grateful to an anonymous referee for his/her detailed and helpful comments.

%-----------------------------------------

 \end{document}